\documentclass[a4paper,10pt]{article}
\usepackage[dvips]{graphicx}
\usepackage{latexsym}
\usepackage{amsmath,amssymb,epsfig,float}

\newcommand{\be}{\begin{equation}}
\newcommand{\ee}{\end{equation}}
\newcommand{\ba}{\begin{array}}
\newcommand{\bqa}{\begin{eqnarray}}
\newcommand{\eqa}{\end{eqnarray}}
\newcommand{\cO}{{\cal O}}

\newcommand{\ket}{\,\rangle}
\newcommand{\bra}{\langle \,}

\begin{document}
\title{ \bf \boldmath Chiral dynamics in $U(3)$ unitary  chiral perturbation theory }
  \author{Zhi-Hui~Guo$^{a,b}$, J.~A.~Oller$^{b}$  and J.~Ruiz de Elvira$^{c}$
  \vspace{0.3cm}
\\ {\footnotesize $^a$ Department of Physics, Hebei Normal University, 050024 Shijiazhuang, P.~R.~China. }
  \\ {\footnotesize $^b$ Departamento de F\'isica, Universidad de Murcia,  E-30071 Murcia, Spain. }
 \\ {\footnotesize $^c$ Departamento de F\'isica Te\'orica II, Universidad Complutense de Madrid, E-28040 Madrid, Spain. }
 }

\date{\today}
\maketitle

\begin{abstract}
We perform a complete one-loop calculation of meson-meson scattering, and of the scalar and pseudoscalar
form factors in $U(3)$ chiral perturbation theory with the inclusion of explicit resonance fields.
 This effective field theory takes into account the low-energy effects of the QCD $U_A(1)$ anomaly explicitly in
the dynamics. The calculations are supplied by non-perturbative unitarization techniques that provide the final results for the
meson-meson scattering partial waves and the scalar form factors considered.
We present thorough analyses on the scattering data, resonance spectroscopy,
spectral functions, Weinberg-like sum rules and semi-local duality.
The last two requirements establish relations between the scalar spectrum with
the pseudoscalar and vector ones, respectively.
The $N_C$ extrapolation of the various quantities is studied as well. The fulfillment of all these non-trivial aspects of the
QCD dynamics by our results  gives a strong support to the emerging picture for the scalar dynamics and its related spectrum.
\end{abstract}


\newpage

Chiral symmetry and $U_A(1)$ anomaly are two prominent features of QCD in the low energy sector.
Chiral perturbation theory ($\chi$PT) \cite{weinberg79,gasser84,gasser85} that exhaustively exploits
 chiral symmetry as well as its spontaneous and explicit breaking to constrain the dynamics allowed,
has proven as a reliable tool to analyze
the QCD low energy processes involving the octet of pseudo-Goldstone bosons $\pi,$ $K$ and $\eta$.
On the other hand, the $U_A(1)$ anomaly of QCD provides a natural explanation of the massive state
$\eta'$~\cite{oriua,ohta}. The consideration of a variable number of colors ($N_C$) in QCD is enlightening.
 An important finding from large $N_C$ QCD \cite{largenc} is that the
$U_A(1)$ anomaly is $1/N_C$ suppressed and thus the $\eta'$ meson becomes the ninth Goldstone boson
at large $N_C$ in the chiral limit~\cite{witten79npb}. This poses strong constraints on the allowed forms of the chiral operators
involving the $ \eta'$ field, which generalizes the conventional
$SU(3)$ $\chi$PT~\cite{gasser85} to the $U(3)$ version~\cite{oriua,herrera97npb,kaiser00epjc}.
Thus  $U(3)$ $\chi$PT is a serious theory to incorporate the $\eta'$ as a dynamical degree of freedom in the chiral effective Lagrangian
approach and hence deserves of detailed calculations. Though the one-loop renormalization and construction
of the corresponding $\cO(p^4)$ Lagrangian are performed in Refs.~\cite{herrera97npb,kaiser00epjc},
further calculations still need to be carried out. Recently the calculation of the one-loop  meson-meson scattering amplitudes
was completed in Ref.~\cite{guo11prd}, and the non-strangeness changing scalar and pseudoscalar form factors are calculated in the present
work.

Based on the calculated scattering amplitudes and form factors from  $U(3)$ $\chi$PT, we then study semi-local duality \cite{collinsbook,pelaez11prd}
between  Regge theory and the hadronic degrees of freedom (h.d.f.) and construct the spectral functions
to investigate the Weinberg-like spectral function sum rules \cite{wsr} among the scalar and pseudoscalar correlators.
The $N_C$ evolution of the resonance poles, semi-local duality and two-point correlators are also studied. In the physical case,
i.e. $N_C=3$, the $f_0(600)$ resonance (also called $\sigma$) plays important roles for the fulfillment of both semi-local
duality and the  Weinberg-like spectral function sum rules. However, according to the study of Ref.~\cite{guo11prd} that employs a similar approach as the one
 used here, when  $N_C$ increases the $f_0(600)$ resonance evolves deeper in the complex
energy plane and barely contributes at large $N_C$. Interestingly, we find that  at large $N_C$ the contribution from the
singlet scalar resonance $S_1$ with a mass around 1~GeV, that is part of the $f_0(980)$ resonance at $N_C=3$,
becomes  more and more important for larger values of  $N_C$. Then, two markedly different pictures for the scalar dynamics emerge
as a function of $N_C$. For the physical case the $f_0(600)$ is the scalar resonance mainly responsible to counterbalance the vector resonance $\rho(770)$ in
 semi-local duality. It also counterbalances the contributions from the octet of scalar resonances,
the nonet of the pseudo-Goldstone bosons and also from the lightest multiplet of pseudoscalar resonances in the  Weinberg-like spectral sum rules.
However, at large $N_C$ the remnant component (a $\bar{q}q$-like one) of the $f_0(980)$ is responsible for the strength in the scalar dynamics.
Though these two pictures differ dramatically they evolve
continuously from one to the other as $N_C$ varies. We present the discussions in more detail next.

In the perturbative calculations, we include the tree level exchanges of resonances explicitly ~\cite{ecker89npb},
instead of considering the local chiral operators from the higher order Lagrangian~\cite{herrera97npb,kaiser00epjc}.
 We then assume tacitly the saturation by resonance exchange of the (next-to-leading) chiral counterterms \cite{ecker89npb}.
The relevant Lagrangian has been presented in detail in Ref.~\cite{guo11prd}. In addition we also include
the exchange of pseudoscalar resonances here, which are absent in \cite{guo11prd}.
Their effects in  meson-meson scattering turn out to be small, but they play a crucial role
to establish the  Weinberg-like spectral sum rules for the difference between the scalar-scalar ($SS$) and pseudoscalar-pseudoscalar ($PP$)
correlators ($SS-PP$).

The pseudoscalar resonance Lagrangian
introduced in \cite{ecker89npb} produces the mixing between the pseudoscalar resonances
and the pseudo-Goldstone bosons. Nevertheless this mixing can be eliminated at the Lagrangian level
through a chiral covariant redefinition of the resonance fields, which results in two local chiral
operators at the $\cO(p^4)$ level~\cite{longpaper}. We remind that the nature of the pseudoscalar resonances
is still a controversial issue and their parameters are not accurately measured yet~\cite{pdg}.
So in order to compensate the uncertainties on the pseudoscalar resonance properties, as well as our simple parameterization here in
terms of simple bare propagators in the spirit of the narrow resonance approach,\footnote{E.g. see Ref.~\cite{albaladejo10prd} for a
refined treatment of the pseudoscalar resonances as dynamically generated resonances from the interactions between the scalar resonances
and the pseudo-Goldstone bosons.}  we include an $L_8$-like operator~\cite{gasser85}.

\begin{figure}[ht]
\begin{center}
\includegraphics[angle=0, width=0.98\textwidth]{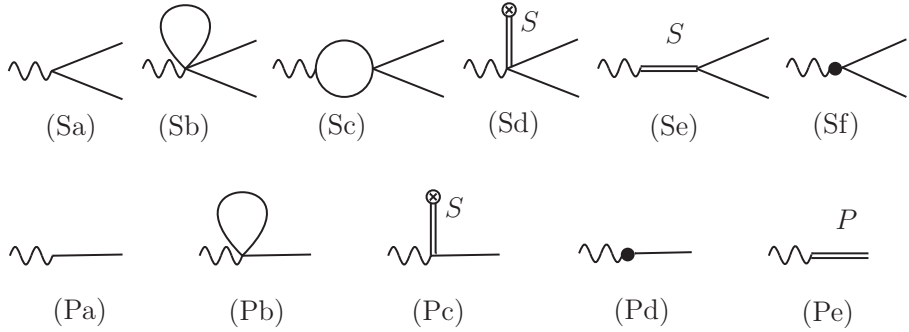}
\caption{{\small Feynman diagrams for the calculations of the scalar (first row) and pseudoscalar (second row) form factors.
 The wavy lines denote either the scalar or the pseudoscalar external source. See the text for more details.
 }}
\label{fig.feynspff}
\end{center}
\end{figure}

We show the pertinent Feynman graphs for the scalar
form factors of the pseudo-Goldstone pairs and the pseudoscalar form factors in the first and second rows of
Fig.~\ref{fig.feynspff}, in order.  The scalar form factor of a pseudo-Goldstone boson pair $PQ$, $F_{PQ}^a(s)$, is defined as
\begin{equation}\label{defsff}
 F_{PQ}^a(s) = \frac{1}{B} \bra 0|S^a|\, PQ \,\ket\,,
\end{equation}
while the pseudoscalar form factor of the pseudoscalar $P$, $H_P^a(s)$, corresponds to
\begin{equation}
\label{defpff}
 H_P^a(s) = \frac{1}{B} \bra 0|P^a|\, P \,\ket\,.
\end{equation}
In the equations above the scalar and pseudoscalar currents are $S^a = \bar{q} \lambda_a q$
and $P^a= i \bar{q} \gamma_5  \lambda_a q$, in order, with $\lambda_a$ the Gell-Mann matrices for $a=1,\ldots,8$ and $\lambda_0=I_{3\times 3}\sqrt{2/3}$ for $a=0$. On the other hand,  $B$ is proportional to the quark condensate in the chiral limit \cite{guo11prd}. In Fig.~\ref{fig.feynspff} the wavy lines correspond to either the scalar or pseudoscalar external sources,
the single straight lines to the pseudo-Goldstone bosons and the double lines to the scalar ($S$) and
pseudoscalar ($P$) resonances. The cross in diagram (Sd) and (Pc)
indicates the coupling between the scalar resonance and the vacuum.
The dot in the diagrams (Sf) and (Pd) corresponds to the vertices involving only pseudo-Goldstone bosons
beyond the leading order. They can stem from many sources, such as from the local terms that originate after removing the mixing between the
pseudo-Goldstone bosons and the pseudoscalar resonances. A detailed account, including explicitly all the relevant expressions, will be presented
in Ref.~\cite{longpaper}.

In $U(3)$ $\chi$PT it is necessary to resum the unitarity loops due to the large $s$-quark mass and the large anomaly mass.
Consequently, the pseudo-Goldstone boson thresholds are much larger than the typical three-momenta in many kinematical regions,
which increases the contributions from the reducible two pseudo-Goldstone boson loops \cite{weinberg91npb}.
Moreover, we are also interested in the resonance energy region where the  unitarity upper bound
in partial wave amplitudes can be easily reached, so that it does not make  sense to treat unitarity perturbatively
as in $\chi$PT for these energy regions. Hence one must resum the unitary cut and we use Unitary $\chi$PT (U$\chi$PT) to accomplish this resummation. This approach is based on the N/D
method \cite{oller99prd}  to resum the unitarity chiral loops both for the partial wave scattering amplitudes
and the form factors. The application of these unitarization techniques to the form factors is discussed in
 Refs.~\cite{oller00prd,meissner01npa,oller05prd}. The partial waves
from $U(3)$ unitary $\chi$PT plus the resonance exchanges at tree level were already discussed in Ref.~\cite{guo11prd}, we
now build the unitarized scalar form factors in a similar fashion \cite{meissner01npa}. Our master equation in matrix notation is
\begin{equation}\label{defunitarizedF}
F^I(s) =  \big[ 1 + N^{IJ}(s) \, g^{IJ}(s) \big]^{-1} R^I(s)\,,
\end{equation}
where
\begin{eqnarray}\label{defNyR}
 R^{I}(s) = {F^{I}(s)}^{\rm (2)+Res+Loop} + T^{IJ}(s)^{\rm (2)} \,g^{IJ}(s)\, F^{I}(s)^{\rm (2)}\,.
\end{eqnarray}
In the previous equation  $T^{IJ}(s)$ is a matrix whose elements are the partial wave scattering amplitudes with definite isospin $I$
and angular momentum $J$. We refer to Ref.~\cite{guo11prd} for details about $T^{IJ}(s)$, $N^{IJ}(s)$ and $g^{IJ}(s)$.
The quantity  ${F^{I}(s)}^{\rm (2)+Res+Loop}$ denotes the scalar form factors of
the Goldstone pairs depicted in the first row of Fig.~\ref{fig.feynspff}. The superscripts (2), Res
and Loop  stand for the perturbative results from the leading order,
resonance contributions and chiral loops, respectively. The vector function $R^I(s)$ in Eq.~\eqref{defNyR} stems from the perturbative calculations of
the form factors and it does not contain any cut singularity \cite{oller00prd,meissner01npa}.

The two-point scalar and pseudoscalar correlators, $\Pi_{S^a}$ and $\Pi_{P^a}$, respectively, are defined as
\begin{equation}
\delta^{ab} \,\Pi_R(p^2) = i \int d^4 x \, e^{i p \cdot x} <0| T [R^a(x) R^b(0)] |0>\,,
\end{equation}
with $R^a=S^a$ or $P^a$.
After the establishment of the unitarized scalar form factors in Eq.~\eqref{defunitarizedF}, we are ready to
calculate the scalar spectral function or the imaginary part of the two-point scalar correlator
\begin{eqnarray} \label{defscspecf}
 {\rm Im} \,\Pi_{S^a} (s) = \sum_{i} \rho_i(s) \left| F_i^a(s) \right|^2\, \theta(s-s_i^{\rm th})\,,
\end{eqnarray}
with $\theta(x)$  the Heaviside step function. The kinetic space factor $ \rho_i(s)$ is defined as
\begin{equation}\label{defkineticsigma}
 \rho_i(s) = \frac{\sqrt{[s-(m_A + m_B)^2][s-(m_A - m_B)^2]}}{16 \pi\, s} \,,
\end{equation}
where $m_A, m_B$ are the masses of the two particles in the $i_{\rm th}$ channel, $s$ is the energy squared in the center of mass frame
and $s_i^{\rm th}=(m_A+m_B)^2$ denotes the threshold.   We focus on the cases with $a=0,$ 3 and 8, which conserve strangeness.
The values $a=0$ and 8 correspond to the isoscalar case $I=0$, and there are five relevant channels:
$\pi\pi$, $K\bar{K}$, $\eta\eta$, $\eta\eta'$ and $\eta'\eta'$. For $a=3$ one has the isovector case $I=1$
and three channels are involved: $\pi\eta$, $K\bar{K}$ and $\pi\eta'$. We adopt the isospin bases and employ the
unitarity normalization as used in Ref.~\cite{guo11prd}.
Another important observable that can be extracted from the scalar form factor is the quadratic pion scalar radius $\bra r^2 \ket_S^\pi$ defined from the Taylor expansion around the origin of the pion scalar form factor as
\begin{equation}\label{defscradius}
F_{\pi\pi}^{\bar{u}u+\bar{d}d}(s) = F_{\pi\pi}^{\bar{u}u+\bar{d}d}(0) \bigg[  1 + \frac{1}{6}\bra r^2 \ket_S^\pi \,s   + ... \bigg]\,,
\end{equation}
with
\begin{eqnarray}\label{defsffuuddss}
m_\pi^2 F_{\pi\pi}^{\bar{u}u+\bar{d}d}(s) \equiv 2B m \bra 0| \bar{u}u+\bar{d}d |\pi\pi\ket
= 2 B m \bigg[ \frac{F_{\pi\pi}^{a=8}(s)}{\sqrt{3}} + \frac{\sqrt{2} \,F_{\pi\pi}^{a=0}(s)}{\sqrt{3}} \bigg]\,,
\end{eqnarray}
where $m$ is the up or down current quark mass (isospin breaking is not considered in this work).

The pseudoscalar spectral function is related to the pseudoscalar form factors, $H_P^a(s)$, depicted
in the second row of Fig.~\ref{fig.feynspff}, by
\begin{equation}\label{defpsspecf}
{\rm Im}\, \Pi_{P^a}(s) = \sum_i \pi\, \delta(s-m_{P_i}^2)\, |H^a_i(s)|^2\,,
\end{equation}
where we do not consider  multiple-particle intermediate states.
In the above equation $\delta(x)$  stands for the Dirac $\delta$ function,
$m_{P_i}$ corresponds to the masses of the pseudo-Goldstone bosons or the pseudoscalar resonances
with the same quantum numbers as the considered spectral function.

Another interesting object that we study is the so-called semi-local (or average) duality in scattering \cite{collinsbook,pelaez11prd}.
We quantify semi-local duality in $\pi\pi$ scattering between the Regge theory and h.d.f.,
by employing the useful ratio between the amplitudes with well-defined $I$ in the $t$-channel,  as proposed in~\cite{pelaez11prd},
\begin{equation}\label{defFratio}
 F_n^{I I'} = \frac{ \int_{\nu_1}^{\nu_{\rm max} } \nu^{-n}\, {\rm Im}\, T_{\rm t}^{(I)}(\nu, t) \,d\nu}
{\int_{\nu_1}^{\nu_{\rm max}} \nu^{-n}\, {\rm Im}\, T_{\rm t}^{(I')}(\nu, t) \,d\nu}\,.
\end{equation}
In this equation the isospin is indicated by the superscript and $\nu = \frac{s-u}{2} = \frac{2s + t -4m_\pi^2}{2}$, with $s,$ $t$ and $u$
the standard Mandelstam variables.  The  relations between the $t$-channel well-defined isospin amplitudes, $T_{\rm t}^{(I)}(s,t)$,
and those with well-defined isospin in the $s$-channel, $T_{\rm s}^{(I)}(s,t)$, are \cite{collinsbook}
\begin{eqnarray}\label{tsrelation}
T_{\rm t}^{(0)}(s,t) &=& \frac{1}{3} T_{\rm s}^{(0)}(s,t) + T_{\rm s}^{(1)}(s,t) + \frac{5}{3} T_{\rm s}^{(2)}(s,t) \,, \nonumber \\
T_{\rm t}^{(1)}(s,t) &=& \frac{1}{3} T_{\rm s}^{(0)}(s,t) + \frac{1}{2} T_{\rm s}^{(1)}(s,t) - \frac{5}{6} T_{\rm s}^{(2)}(s,t) \,, \nonumber \\
T_{\rm t}^{(2)}(s,t) &=& \frac{1}{3} T_{\rm s}^{(0)}(s,t) - \frac{1}{2} T_{\rm s}^{(1)}(s,t) + \frac{1}{6} T_{\rm s}^{(2)}(s,t) \,.
\end{eqnarray}
Since Regge exchange is highly suppressed for the exotic $I=2$ case in the $t$-channel, Regge theory predicts a
vanishing value for the ratios $F_n^{21}$ and $F_n^{20}$. In the following we shall focus on the ratio $F_n^{21}$ to test semi-local duality in
order to make a close comparison with Ref.~\cite{pelaez11prd}.
We study the scattering for two values of $t$, $t=0$ (forward scattering) and  $t=4m_\pi^2$, in order to test the stability of the results for
different small values of $t$ compared with GeV$^2$. The lower integration limit  $\nu_1$ is always set to
the threshold point and we concentrate on the energy region with $\nu_{\rm max}=2$~GeV$^2$ for the ratio in Eq.~\eqref{defFratio}.
To calculate in Eq.~\eqref{tsrelation} the imaginary parts of the $t$-channel well-defined isospin amplitudes, ${\rm Im}\,T_{\rm t}^{(I)}(s,t)$,
 we  need to know ${\rm Im}\,T_{\rm s}^{(I)}(s,t)$, which can be decomposed in the center of mass frame in a partial wave expansion as
\begin{equation}\label{pwdecompose}
  {\rm Im}\, T_{\rm s}^{(I)}(\nu, t) = \sum_J (2 J + 1)\,  {\rm Im}\, T^{IJ}(s) \, P_J(z_s)\,,
\end{equation}
with $z_s = 1 + 2 t /(s - 4m_\pi^2)$, the cosine of the scattering angle, and $P_J(z_s)$ the Legendre polynomials.
The partial waves $T^{IJ}(s)$ were already carefully studied in Ref.~\cite{guo11prd} within $U(3)$ unitary $\chi$PT,
and we extend the results there by including
the contributions from the exchange of the pseudoscalar resonances.

We point out that all the parameters entering the form factors also appear in the unitarized scattering amplitudes and
in the expressions for the masses
of the pseudo-Goldstone bosons. Hence, once the  unknown parameters are determined by the fit to scattering data
and the pseudo-Goldstone masses, we can completely predict the form factors and spectral functions.
 By using the best fit in Eq.~(55) of Ref.\cite{guo11prd} for the calculation of the pion scalar form factor, a small quadratic pion scalar radius is obtained
$\bra r^2 \ket_S^\pi = 0.43 \,{\rm fm}^2$, which is around 30\% less than the dispersive
result $0.61 \,{\rm fm}^2$ in \cite{colangelo01npb}. One way to improve the pion scalar radius is to increase the value
of $L_5$~\cite{gasser85}. It is found  in Ref.~\cite{jamin02npb} that a second multiplet of scalar resonances around 2~GeV
contributes around 50\% of $L_5$. Thus, we shall include this second scalar
multiplet in our analysis and we take the values for its resonance parameters from the preferred fit Eq.~(6.10) of  Ref.~\cite{jamin02npb}.
 The inclusion of this second scalar nonet and of the pseudoscalar resonance exchanges  requires to perform a new fit.
The resulting quality of the new fit and also the resonance spectroscopy, which will be given in detail in Ref.~\cite{longpaper},
are quite similar to the ones of Ref.\cite{guo11prd}, so we refrain from discussing them further here. But the new fit improves the pion scalar
radius to $0.49^{+0.01}_{-0.03}$~fm$^2$, being around a 14\% larger than the result from the best fit of Ref.~\cite{guo11prd}.

\begin{figure}[H]
\includegraphics[angle=0, width=0.98\textwidth]{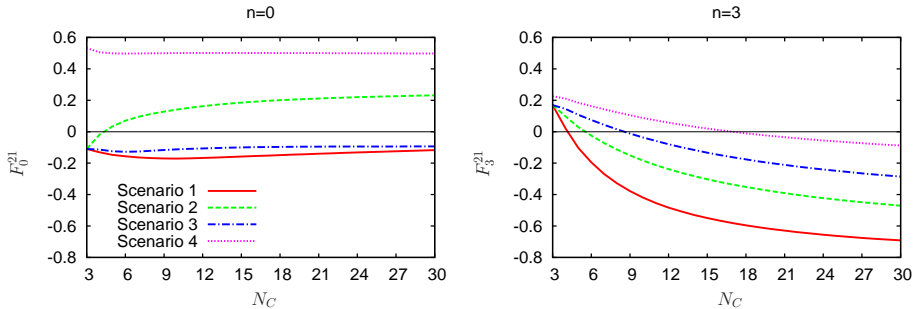}
\caption{{\small Evolution of $F_n^{21}(t=4m_\pi^2)$ from $N_C=3$ to 30 for the four scenarios considered. See the text and Table~\ref{tab:scenarios} for the
meaning of each scenario. We verify that the ratios evaluated at $t=0$ are similar. }}
\label{fig.ncdualityf21}
\end{figure}

Let us consider other interesting consequences of the new fit. As we commented previously, an important advantage of
$U(3)$ $\chi$PT, compared with the $SU(2)$ or $SU(3)$ versions, is that it incorporates the singlet $\eta_1$ that becomes the ninth
Goldstone boson at large $N_C$ in the chiral limit and thus $U(3)$ $\chi$PT is more adequate to discuss the large $N_C$ dynamics.
The leading order $N_C$ scaling for the various parameters in our theory was already given in \cite{guo11prd}.
For the pion decay constant $F_\pi$, we always take both the leading and sub-leading $N_C$ terms which were calculated
in Ref.~\cite{guo11prd} at the one-loop level in $U(3)$ $\chi$PT. In addition to only including the leading
$N_C$ behavior for the remaining parameters, referred as Scenario 1,
we also consider other three scenarios that include  sub-leading $N_C$ scaling for the resonance parameters.
Through the fit to experimental data, we determine the values of the parameters at $N_C=3$.
By imposing short distance constraints, the resonance parameters that then result at large $N_C$ are already discussed in many
contexts~\cite{ecker89plb,pich11jhep,guo10prd,guo07jhep}. Among these constraints, we take the one from
the vector resonance sector, which should be quite reliable
 due to the well established $\rho(770)$ $\bar{q}q$-like resonance at large $N_C$.
 An updated version of the constraint on $G_V$, a coupling describing
the vertices of the $\rho(770)$ with pions, is revealed in many recent works \cite{pich11jhep,guo10prd,guo07jhep,guo11prd} as
\begin{equation}\label{gvatnc}
G_V = \frac{F}{\sqrt{3}}\,,
\end{equation}
with $F$ the pion decay constant at large $N_C$. The extrapolation function for $G_V$ is uniquely
fixed if one considers contributions up to and including next-to-leading order in the large $N_C$ expansion and requires $G_V$ to
take the value given by the fit at $N_C=3$ and the result in Eq.~\eqref{gvatnc} at large $N_C$.
We present the detailed expressions  in Ref.~\cite{longpaper}. We refer the situation including
the sub-leading piece for $G_V$ as Scenario 2. In Scenario 3, on top of the setups in Scenario 2, we impose that
$M_\rho$ and $M_{S_1}$ approach to the same value at large $N_C$, which can be realized naturally by tuning
the corresponding parameters at the level of 16\% from the values at $N_C=3$.
While in Scenario 4, we keep all the constraints from Scenario 3 and include the tensor resonances,
which are the dominant contributions to the $D$-wave amplitudes.
We follow Ref.~\cite{ecker07epjc} to include the tensor resonances in meson-meson scattering and also
take the numerical value for the tensor coupling as determined there. The explicit calculation will be also given in detail in Ref.~\cite{longpaper}.
The characteristics of the different scenarios considered are summarized in Table~\ref{tab:scenarios}.
As proposed in Ref.~\cite{pelaez11prd}, $F^{21}_n$ with $n=0,$ 1, 2 and 3 are the relevant ratios in our considered energy region. We show the $N_C$ evolution of the ratio $F^{21}_n$ from Eq.~\eqref{defFratio} in Fig.~\ref{fig.ncdualityf21} for $n=0$ and 3.
And more details for $n=1$ and 2 will be given in Ref.~\cite{longpaper}.

\begin{table}[ht]
\begin{center}
\begin{tabular}{|p{1.9cm}|p{1.5cm}| p{1.5cm}| p{1.5cm} | }
\hline
 & $G_V$ & $M_\rho$, $ M_{S_1}$ &  $D$-wave
\\
\hline
Scenario 1 & $-$ & $-$ & $-$
\\
\hline
Scenario 2 & $\surd$ & $-$ & $-$
\\
\hline
Scenario 3 & $\surd$ & $\surd$ & $-$
\\
\hline
Scenario 4 & $\surd$ & $\surd$ & $\surd$
\\
\hline
\end{tabular}
 \caption{{\small Description of Scenarios 1--4. In the second and third columns the symbol $\surd$ ($-$)
denotes that the sub-leading $N_C$ scaling for the corresponding parameters is (not) considered.
In the last column, the symbol $\surd$ ($-$) means that we do (not) consider the contribution from the $D$-waves.
 } }\label{tab:scenarios}
\end{center}
\end{table}

Notice that if the required cancellations between the $I=0$ and $I=1$ partial wave amplitudes in Eq.~\eqref{tsrelation}
did not take place for $T_{\rm t}^{(2)}(s,t)$, as they are required by Regge exchange theory,
the natural value for $\left| F_n^{21} \right|$ would be around 1. While if the semi-local duality is satisfied,
$\left| F_n^{21} \right|$ should approach to zero. So we conclude that Scenario 3 is the best one of the four situations.
The main problem in Scenario 4 is that the tensor resonances give too large contributions
and overbalance the $\rho(770)$ resonance for $n=0$. This seems to indicate that once the tensor resonances are included,  heavier vector
resonances are needed so as to fulfill better semi-local duality for $n=0$.
 A remarkably valuable information that we can get from the study of semi-local duality is its capacity to  distinguish clearly between the different scenarios proposed and hence it provides a tight constraint
on the $N_C$ evolution of the resonance parameters.
In the following  we shall only focus on the $N_C$ running within Scenario 3, since it is the one that satisfies best semi-local duality.

\begin{figure}[H]
\begin{center}
\includegraphics[angle=0, width=0.9\textwidth]{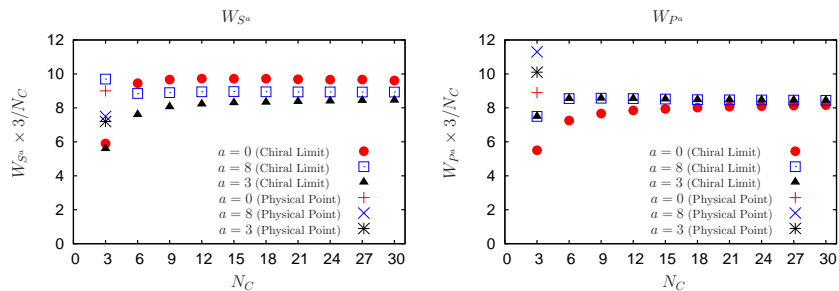}
\caption{{\small $W_i\times 3/N_C$ as a function of $N_C$ within Scenario 3. All of the results are calculated by setting
the upper limit of the integral in Eq.~\eqref{defwi} to $s_0= 3$~GeV$^2$. We check
that the results with $s_0=2.5$~GeV$^2$ and $s_0=3.5$~GeV$^2$ are quite similar.    }}
\label{fig.winc}
\end{center}
\end{figure}

Now, we study the  Weinberg-like spectral sum rules in the scalar and pseudoscalar sectors, which are given by
\begin{equation}\label{defweinbergsr2}
\int_0^{s_0} \big[ {\rm Im}\,\Pi_R(s) - {\rm Im}\,\Pi_{R'}(s) \big]\,  d s
+ \int_{s_0}^{\infty} \big[ {\rm Im}\,\Pi_R(s) - {\rm Im}\,\Pi_{R'}(s) \big]\,d s  = 0\,,
\end{equation}
where $R$, $R'$=$S^a$ or $P^a$, with $a=0,8,3$. With a proper choice of $s_0$, we can calculate the first integral employing the results from the present
 study in the non-perturbative region
and use the results from the operator product expansion (OPE) to calculate the second one.
 According to the OPE study of Ref.~\cite{jamin92zpc} the different spectral functions considered here are equal in the asymptotic region
in the chiral limit.\footnote{The calculation in Ref.~\cite{jamin92zpc} is done up to ${\cal O}(\alpha_s)$
and including up to dimension 5 operators.}
As a result the second integral in Eq.~\eqref{defweinbergsr2} is zero and to test how well the  Weinberg-like spectral function sum rules hold
 reduces to the evaluation of the first integral in Eq.~\eqref{defweinbergsr2}  in the energy region below $\sqrt{s_0}$.
 The relevant spectral functions ${\rm Im}\,\Pi_R$
are calculated through Eq.~\eqref{defscspecf} for the scalar case and from Eq.~\eqref{defpsspecf} for the pseudoscalar one.
To study the dependences of the first integral in Eq.~\eqref{defweinbergsr2} with $s_0$, we try three values
of $s_0$, namely,  $s_0=2.5$, 3.0 and 3.5 GeV$^2$ and we confirm that the results  are quite stable for the different values taken.
In order to display the results in a more compact way, we show the value of the integral
separately for each spectral function
\begin{eqnarray} \label{defwi}
 W_i &=& 16\pi \int_0^{s_0}   {\rm Im}\,\Pi_i(s)  \,  d s\,, \quad  i=S^8,~S^0,~S^3,~P^0,~P^8,~P^3\,,
\end{eqnarray}
instead of the differences between the various correlators.
We show the results for $W_i\times 3/N_C$ in Fig.~\ref{fig.winc} at the physical point and also their $N_C$ evolution in the chiral limit.
In order to study $W_i$ in the chiral limit, we need to perform the chiral extrapolation. Though the resonance parameters are independent
on the quark masses, the subtraction constants introduced through the unitarization procedure depend on them. Indeed it is shown in Ref.~\cite{Jido:2003cb} that in the $SU(3)$ limit case (as in the chiral limit) all of them should be the same for any $PQ$ pair involving the $\pi$, $K$ and $\eta_8$ pseudoscalars. Indeed, we find that in the chiral limit there exists a reasonable region for a common value of all the subtraction constants  where the values of the two-point correlators are stable and  Weinberg sum rules are improved comparing with the physical situation. This region includes values similar to the ones fitted.  In Fig.~\ref{fig.winc}, we show the typical result in this region and
normalize by the factor $3/N_C$ because $W_i$ scales as $N_C$, as it is also clear from the results plotted in the figure.
Focusing on the points at the chiral limit case in Fig.~\ref{fig.winc}, the relative variance among the six numbers,
i.e. the square root of the variance divided by their mean value \cite{longpaper}, is found to be 10\%, implying that  the
Weinberg-like spectral function sum rules in the $SS-SS$, $PP-PP$ and $SS-PP$ sectors hold quite accurately.
The fulfillment of these sum rules even improves  at large $N_C$ and the relative variance reduces to 5\% for $N_C=30$.

\begin{figure}[H]
\begin{center}
\includegraphics[angle=0, width=0.9\textwidth]{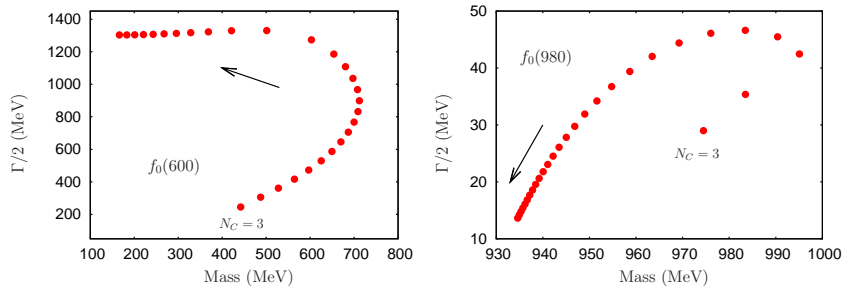}
\caption{{\small Pole trajectories as a function of $N_C$  for the resonances $f_0(600)$ and $f_0(980)$.
We show the results from $N_C=3$ to 30 in one unit step.}}
\label{fig.ncpole}
\end{center}
\end{figure}

Up to now, we have shown that our formalism can simultaneously fulfill  semi-local duality between
the Regge theory and h.d.f. and the Weinberg-like spectral function  sum rules both for the physical case
and large values of $N_C$. Of course this success is based on the fact that we properly take the $N_C$ scaling
for the resonance parameters dictated by the short distance constraint. It is interesting to de-construct the ratio $F_n^{21}$ and
the Weinberg-like spectral sum rules to see how different resonances contribute to them.
At the physical case, we obtain the spectroscopy for various resonances,
such as $f_0(600)$, $f_0(980)$, $f_0(1370)$, $a_0(980)$, $a_0(1450)$, $K^*_0(800)$ (also called $\kappa$), $K^*_0(1430)$,
$\rho(770)$, $K^*(892)$ and $\phi(1020)$, and they agree quite well with their properties reported in the PDG ~\cite{pdg}.
Taking $F_3^{21}$ as an example, we observe an interesting interplay between
the $f_0(600)$ and $f_0(980)$ resonances in the $N_C$ evolution. In Fig.~\ref{fig.ncpole}, we show the $N_C$ trajectories for the $f_0(600)$ and $f_0(980)$, from left to right, respectively. More details about the other resonances will be displayed elsewhere \cite{longpaper}.
For the physical situation, both $f_0(600)$ and $\rho(770)$ give important contributions to $F_3^{21}$,
which leads to a significant cancellation between each other, that is necessary in order to guarantee semi-local duality. While
the $f_0(980)$ only plays a marginal role. But when $N_C$ increases, the $f_0(600)$ pole as shown in Fig.~\ref{fig.ncpole} and in Ref.~\cite{guo11prd},
blows up in the complex energy plane and does not play any significant role at large $N_C$.
In contrast, the $\rho(770)$ resonance falls down to the real axis \cite{pelaezprl1,guo11prd}, behaving as a standard $\bar{q}q$-like
resonance at large $N_C$, and definitely contributes to the ratio $F_3^{21}$.
The scalar strength to cancel the contribution from the $\rho(770)$ comes now from the $f_0(980)$ resonance,
which gradually evolves to the singlet scalar $\bar{q}q$-like $S_1$ when
increasing $N_C$.

It is also worth comparing our results with those from the previous works \cite{pelaez11prd,pelaezprl1,pelaezprl2} based on the use of the Inverse Amplitude Method \cite{dobado}.
The $N_C$ trajectories shown in Fig.\ref{fig.ncpole}, confirm again the results
obtained in \cite{pelaezprl1,pelaezprl2} which predict a non-dominant $\bar
qq$ behavior for the $f_0(600)$. The latter was explained in terms of different
kind of resonances in Ref.~\cite{LlanesEstrada:2011kz}. Note that the
$f_0(600)$ behavior in Fig.~\ref{fig.ncpole}, moving towards lower masses and larger widths, was
found in Refs.~\cite{pelaez11prd,Pelaez:2005fd} by varying the renormalization scale where
the $N_C$ scaling of the $\chi$PT low energy constants applies.
Let us remark that, as it happens in Ref.~\cite{pelaez11prd}, in order to satisfy semi-local duality, we also need a $\bar qq$ component
around 1 GeV. However, this work presents an alternative to Refs.~\cite{pelaezprl2,pelaez11prd} because at $N_C=3$
such a $\bar qq$ component would belong to the $f_0(980)$ instead to the $f_0(600)$.

 Large cancellations are also required to satisfy the Weinberg-like spectral function sum rules.
For the physical case, the singlet correlator $W_{S^0}$ receives important contributions both from
the $f_0(600)$ and $f_0(980)$.  The octet $W_{S^8}$ mainly gets contribution from the $f_0(1370)$ resonance
and  is also slightly contributed by the $f_0(600)$ and $f_0(980)$.
For $W_{S^3}$, the $a_0(980)$ peak  dominates its spectral function, though it receives
non-negligible contributions from the $a_0(1450)$. However at large $N_C$, the $a_0(980)$ resonance
goes deep in the complex energy plane, like the $f_0(600)$ for the isoscalar case, and hence it does
not contribute to $W_{S^3}$ any more. Instead, the $a_0(1450)$ becomes more important when increasing $N_C$
and finally matches the contributions from the $f_0(980)$ in the singlet correlator $W_{S^0}$ and $f_0(1370)$ in
$W_{S^8}$, so that the Weinberg-like spectral function sum rules at large $N_C$ are  well satisfied.

Finally, we summarize briefly our work. We perform a complete one-loop calculation of
the scalar and pseudoscalar form factors within  $U(3)$ unitary $\chi$PT, including the tree-level exchange of resonances.
The spectral functions of the two-point correlators are constructed by using the resulting form factors (which are unitarized for the case of the
scalar ones).
After updating the fit in Ref.~\cite{guo11prd}, which is also extended by including the explicit exchange of pseudoscalar resonances,
we study the resonance spectroscopy, quadratic pion scalar radius, and the fulfillment of semi-local duality
and the Weinberg-like spectral function sum rules in the $SS-SS$, $PP-PP$ and $SS-PP$ cases, which are well satisfied.
We show that it is important to take under consideration the high energy constraint for $G_V$, Eq.~\eqref{gvatnc}, in order to keep
semi-local duality when varying $N_C$. An interesting interplay between different resonances when studying the $N_C$
evolution of semi-local duality and the Weinberg-like spectral sum rules is revealed. In the former case the scalar and
vector spectra appear  tightly related and in the latter one the same can be stated for the scalar and pseudoscalar ones.

The idea to study the Weinberg sum rules in $U(3)$ $\chi$PT was brought up by our colleague
J.~Prades, who unfortunately passed away.
We would like to express our gratitude to his help in this subject.
We also acknowledge the valuable discussions with J.~R.~Pel\'aez.
This work is partially funded by the grants MEC  FPA2010-17806, the Fundaci\'on S\'eneca 11871/PI/09,
the BMBF grant 06BN411, the EU-Research Infrastructure
Integrating Activity ``Study of Strongly Interacting Matter" (HadronPhysics2, grant No. 227431)
under the Seventh Framework Program of EU and the Consolider-Ingenio 2010 Programme CPAN (CSD2007-00042).
Z.H.G. also acknowledges the grants National Natural Science Foundation of China (NSFC) under contract No. 11105038,
Natural Science Foundation of Hebei Province with contract No. A2011205093 and Doctor Foundation of Hebei Normal
University with contract No. L2010B04.

\end{document}